\documentclass[12pt]{iopart}

\usepackage[cp1251]{inputenc}
\usepackage{iopams}
\usepackage{graphicx,epsfig}
\begin{document}

\title[Asymptotic superselection rule]{Asymptotes of non-stationary solutions of the Schr\"{o}dinger equation for a
particle interacting with a one-dimensional $\delta$-potential}

\author{N. L. Chuprikov}

\address{Tomsk State Pedagogical University, 634041, Tomsk, Russia}
\ead{chnl@tspu.edu.ru} \vspace{10pt}

\begin{abstract}
It is shown that non-stationary solutions of the Schr\"{o}dinger equation, which describes the quantum dynamics of a
particle in the field of a one-dimensional delta potential (1DDP), are divided into two classes: some define pure states
that have no free dynamics as $t\to\mp\infty$; others define states with asymptotically free dynamics but represent mixed
states in whose space the asymptotic superselection rule holds. That is, according to the Schr\"{o}dinger equation, {\it
pure scattering} states predicted by the conventional model of this scattering process do not exist. On mixed scattering
states, the Hamiltonian with 1DDP is defined only in superselection sectors. The scattering process with one-sided
incidence of a particle on 1DDP represents a decoherence process in a closed system.
\end{abstract}

\noindent PACS: 03.65.$-$w; 03.65.Ta; 03.65.Nk; 03.65.Xp

\newcommand{\ppp}{\mbox{\hspace{5mm}}}
\newcommand{\ppa}{\mbox{\hspace{15mm}}}
\newcommand{\ppb}{\mbox{\hspace{20mm}}}
\newcommand{\ooo}{\mbox{\hspace{3mm}}}
\newcommand{\ooa}{\mbox{\hspace{1mm}}}

\section{Introduction} \label{int}

\hspace*{\parindent} Exactly one hundred years have passed since the Schr\"{o}dinger equation was published. It is
currently considered the fundamental equation of nonrelativistic quantum mechanics, adequately describing single-particle
processes occurring in the microworld, as all known quantum mechanical models based on solutions of this equation are
completely consistent with experiment.

However, in modern quantum mechanics, the Schr\"{o}dinger equation is used not only to construct models of specific
quantum processes. For purely intuitive reasons, it is also considered to justify the superposition principle in any
single-particle quantum process in the (simply connected) space $\mathbb{R}^n$ ($n=1,2,3$) described by this equation; any
solution of this equation (normalized to unity) is considered one of the possible states of the particle participating in
this process.

Dirac says that ``{\it The assumption of superposition relationships between the states leads to a mathematical theory in
which the equations that define a state are linear in the unknowns.}'' Messiah, regarding the superposition principle and
the Schr\"{o}dinger equation, which the wave function $\Psi$ must satisfy, specifies on p. 69 in \cite{Mess}: ``{\it The
equation must be linear and homogeneous; then the wave [wave function] satisfies the superposition principle,
characteristic of wave processes in the general case. Namely, if $\Psi_1$ and $\Psi_2$ are solutions of the equation, then
any linear combination $\lambda_1\Psi_1+\lambda_2\Psi_2$ of these functions is a solution of the same equation}'' (see p.
10 in \cite{Dir}).

According to existing formulations of the superposition principle for a spinless particle (as a closed system), all states
of a particle whose quantum dynamics is described by the Schr\"{o}dinger equation should be considered as {\it pure}
quantum states. This means (see p. 108 in \cite{Jau}) that the (linear self-adjoint) operators of observables must act
{\it irreducibly} in the space $\mathcal{H}$ formed by the states of this system; $\mathcal{H}$ should not contain
nontrivial subspaces (different from $\mathcal{H}$) that would be invariant with respect to the action of these operators.
This requirement, it is sometimes embedded in the formulation of the superposition principle (see, e.g., p. 262 in
\cite{Jae}): ``Each physical system is represented by a Hilbert space and described by physical quantities, and a state is
represented by linear operators in that space''.

However, the unconditional application of the superposition principle is justified only in problems with an external field
$V(\vec{r})$ such that the probability of detecting a particle at infinity is zero. In this case, its energy spectrum is
discrete, and any of its non-stationary states can be represented as a linear combination of a countable set of basis
functions—the eigenfunctions of the energy operator. It is important to emphasize that all known models of the quantum
dynamics of a spinless, non-relativistic particle were just developed for this class of potentials $V(\vec{r})$. These
models are exact in the sense that they extract information about the processes under study solely from the
Schr\"{o}dinger equation, without resorting to any additional assumptions regarding their properties.

A completely different situation arises in scattering problems, where it is important to study the particle dynamics as
$t\to\mp\infty$ (in these problems, the probability of finding a particle at infinity is non-zero). Currently, there is no
model (even for a particle scattering on one-dimensional short-range potential barriers $V(x)$) in which information about
the quantum dynamics of a particle as $t\to\mp\infty$ would be extracted from non-stationary solutions of the
Schr\"{o}dinger equation. According to modern quantum scattering theory, this is simply not necessary, since all the
necessary information about the properties of the asymptotes of scattering states is embedded in the concept of wave
operators, and therefore the main question of scattering theory is to prove the existence and completeness of wave
operators (see, e.g., \cite{Dol,Ens,Alb,Ford}).

But the point is that the concept of wave operators is based on assumptions whose consistency with the Schr\"{o}dinger
equation has never been tested:
\begin{itemize}
\item[(a)] the convergence (in norm) of the state of a particle, as $t\to\mp\infty$, to the solution of the Schr\"{o}dinger equation for
a free particle means that this state describes free dynamics at $t\to\mp\infty$ and, therefore, is a scattering state;
\item[(b)] the Hamiltonian with any short-range potential is a linear operator on the scattering states, and the scattering
states themselves are pure quantum states.
\end{itemize}
In other words, the modern quantum theory of particle scattering by short-range potentials is based on the assumption that
all non-stationary states from the subspace $\mathcal{H}_{ac}$ corresponding to the absolutely continuous spectrum of the
Hamiltonian operator are scattering states, and all these states (as solutions of the linear Schr\"{o}dinger equation,
which (as is assumed) guarantees the fulfillment of the superposition principle) are pure states. And it is important to
emphasize once again that the consistency of this assumption with solutions of the Schr\"{o}dinger equation, even for the
simplest one-dimensional potential $V(x)$, has never been verified.

However, the proposed model of scattering a particle on a one-dimensional $\delta$-potential, which relies exclusively on
solutions of the Schr\"{o}dinger equation for this potential, serves as a counterexample to the quantum scattering theory
presented in \cite{Dol,Ens,Alb,Ford}. An analysis of the asymptotes of the non-stationary solutions of this equation shows
that the assumptions underlying the concept of wave operators are incompatible with each other in this scattering problem:
\begin{itemize}
\item On the one hand, although $\mathcal{H}_{ac}$ does contain pure states that converge in norm as $t\to\mp\infty$ to the
solution of the Schr\"{o}dinger equation for a free particle, they do not describe the free dynamics at $t\to\mp\infty$,
since such states cannot be separated, when $t\to\mp\infty$, into transmitted and reflected wave packets (see Sections
\ref{asymptote} and \ref{invar});
\item On the other hand, there are states in $\mathcal{H}_{ac}$ that describe free dynamics at $t\to\mp\infty$ and are therefore
scattering states, but they are mixed states; these states form a subspace of $\mathcal{H}_{ac}$ in which the asymptotic
superselection rule holds (see sections \ref{asymptote}, \ref{invar}, and \ref{rule}); the Hamiltonian with $\delta$-potential
is defined as a linear operator only in the superselection sectors of this subspace.
\end{itemize}

\section{States of a particle in the field of a one-dimensional $\delta$-potential} \label{start}

Let $V(x)=W\delta(x)$, where $W>0$. In the region ${\mathbb{R}}_-\oplus {\mathbb{R}}_+=\mathbb{R}\setminus \{0\}$, where
${\mathbb{R}}_-=(-\infty,0)$ and ${\mathbb{R}}_+=(0,\infty)$, the wave function $\Psi(x,k)$, describing the stationary
state of the particle, must satisfy the Schr\"{o}dinger equation for a free particle:
\begin{eqnarray} \label{1}
\fl \hat{H}_0\Psi(x,k)\equiv -\frac{\hbar^2}{2m}\frac{d^2\Psi(x,k)}{dx^2}=E(k)\Psi(x,k),
\end{eqnarray}
where $E(k)=\hbar^2k^2/(2m)$ is the particle energy; $k$ is the wave number, and at the point $x=0$ it must satisfy the
boundary conditions
\begin{eqnarray} \label{2}
\fl \Psi(0 +,k)=\Psi(0-,k)=\Psi(0,k),\ppp \Psi^\prime (0+,k)-\Psi^\prime(0-,k)=2\kappa\Psi(0,k);
\end{eqnarray}
$\Psi(0\pm,k)=\lim_{\epsilon\to 0}\Psi(0\pm \epsilon,k)$, and $\Psi^\prime(0\pm, k)=\lim_{\epsilon\to 0}\Psi^\prime(0\pm
\epsilon,k)$; the prime denotes the derivative with respect to $x$; $\kappa=mW/\hbar^2$.

All eigenvalues of the operator $\hat{H}=\hat{H}_0+W\delta(x)$ are doubly degenerate and lie in the region $E>0$ (since
$W>0$, there is no bound state). Two linearly independent solutions for each value of $E(k)$ can be written as
\begin{eqnarray} \label{3}
\fl \Psi_{L}(x,k)= \theta(-x)[e^{ikx}+\mathcal{R}(k)e^{-ikx}]+ \theta(x) \mathcal{T}(k)e^{ikx},\\
\fl \Psi_{R}(x,k)=\theta(-x)\mathcal{T}(k)e^{-ikx}+\theta(x) [e^{-ikx}+\mathcal{R}(k)e^{ikx}]=\Psi_{L}(-x,k); \nonumber
\end{eqnarray}
where $k>0$, and $\theta(x)$ is the Heaviside function ($\theta(0)=1/2$);
\begin{eqnarray} \label{105}
\fl \mathcal{T}(k)=\frac{k}{k+i\kappa}=\mathcal{T}^*(-k);\ooa
\mathcal{R}(k)=-\frac{i\kappa}{k+i\kappa}=\mathcal{R}^*(-k);\ooa \mathcal{T}(k)\mathcal{R}^*(k)
=-\mathcal{T}^*(k)\mathcal{R}(k);
\end{eqnarray}
$|\mathcal{T}(k)|^2$ and $|\mathcal{R}(k)|^2$ are the transmission and reflection coefficients. According to the
formulation of the stationary scattering problem, the function $\Psi_{L}(x,k)$ describes the scattering of a particle with
momentum $\hbar k>0$ incident on the barrier from the left, and the function $\Psi_{R}(x,k)$ describes the scattering of a
particle with momentum $-\hbar k<0$ incident on the barrier from the right.

At first glance, when the stationary solutions are known, it is easy to write in general form the solutions $\Psi_L(x,t)$
and $\Psi_R(x,t)$, which determine (up to a phase factor) the non-stationary scattering states in problems with left- and
right-handed particle incidence on a barrier. However, in reality, a contradictory situation arises here. On the one hand,
according to assumptions (a) and (b), any non-stationary solution constructed on the basis of stationary solutions
(\ref{3}) represents a scattering state. These assumptions require that the solutions $\Psi_L(x,t)$ and $\Psi_R(x,t)$ be
written in the most general form:
\begin{eqnarray} \label{4a}
\fl \Psi_L(x,t)=\frac{1}{\sqrt{2\pi}}\int_{-\infty}^\infty A_L(k,t)\Psi_L(x,k)dk,\ooo A_L(k,t)=\mathcal{A}_L(k)
e^{-iE(k)t/\hbar},
\end{eqnarray}
\begin{eqnarray} \label{4b}
\fl \Psi_R(x,t)=\frac{1}{\sqrt{2\pi}}\int_{-\infty}^\infty A_R(k,t)\Psi_R(x,k)dk,\ooo A_R(k,t)=\mathcal{A}_R(k)
e^{-iE(k)t/\hbar},
\end{eqnarray}
where $\mathcal{A}_L(k)$ and $\mathcal{A}_R(k)$ from $L^2(\widehat{\mathbb{R}})$ are such that the norm of each of the
solutions $\Psi_L(x,t)$ and $\Psi_R(x,t)$ is equal to one; here $\widehat{\mathbb{R}}$ is the $k$-axis. On the other hand,
since the stationary solutions (\ref{3}) are defined only for $k>0$, the functions $\mathcal{A}_L(k)$ and
$\mathcal{A}_R(k)$ in Exps. (\ref{4a}) and (\ref{4b}) must be identically equal to zero for $k\leq 0$.

To see the difference between the non-stationary solutions in these two cases, it is necessary to study the dependence of
the properties of the asymptotes of the solutions $(\ref{4a})$ and $(\ref{4b})$ as $t\to\mp\infty$ on the type of the
functional space to which the functions $\mathcal{A}_L(k)$ and $\mathcal{A}_R(k)$ belong. For this purpose, it is
convenient to switch to the $k$-representation, having first represented the solutions $\Psi_L(x,t)$ and $\Psi_R(x,t)$
taking into account the equality $\theta(x)=[1+\mathrm{sgn}(x)]/2$:
\begin{eqnarray} \label{100}
\fl \Psi_{L,R}(x,t)= \Phi_{L,R}^{inc}(x,t)+\frac{1}{2}\left[f_{L,R}(x,t)+f_{L,R}(-x,t)\right]+
\frac{\mathrm{sgn}(x)}{2}\left[f_{L,R}(x,t)-f_{L,R}(-x,t)\right]; \nonumber \\ \fl f_{L,R}(x,t)=
\frac{1}{\sqrt{2\pi}}\int_{-\infty}^\infty A_{L,R}(k,t) \mathcal{R}(k)e^{ikx}dk.
\end{eqnarray}
Let us introduce wave packets that we consider as candidates for the role of in- and out-asymptotes of non-stationary
solutions $\Psi_{L}(x,t)$ and $\Psi_{R}(x,t)$:
\begin{eqnarray} \label{5a}
\fl \Phi_L^{inc}(x,t)=\frac{1}{\sqrt{2\pi}}\int_{-\infty}^\infty A_L(k,t) e^{ikx}dk,\ppp\ppp\ooa
\Phi_R^{inc}(x,t)=\frac{1}{\sqrt{2\pi}}\int_{-\infty}^\infty A_R(k,t) e^{-ikx}dk, \\ \fl
\Phi_L^{tr}(x,t)=\frac{1}{\sqrt{2\pi}}\int_{-\infty}^\infty A_L(k,t) \mathcal{T}(k)e^{ikx}dk,\ooo\ooa
\Phi_R^{tr}(x,t)=\frac{1}{\sqrt{2\pi}}\int_{-\infty}^\infty A_R(k,t) \mathcal{T}(k)e^{-ikx}dk,\nonumber\\
\fl \Phi_L^{ref}(x,t)=f_{L}(-x,t),\ppp \Phi_R^{ref}(x,t)=f_{R}(x,t).\nonumber
\end{eqnarray}
We need to figure out what class the functions $\mathcal{A}_L(k)$ and $\mathcal{A}_R(k)$ must belong to for these wave
packets to actually play this role; note that $\langle \Phi_{L,R}^{inc}|\Phi_{L,R}^{inc}\rangle= \langle
\Phi_{L,R}^{tr}|\Phi_{L,R}^{tr}\rangle+\langle \Phi_{L,R}^{ref}|\Phi_{L,R}^{ref}\rangle$.

In the $k$-representation, the solutions $\Psi_{L}(x,t)$ and $\Psi_{R}(x,t)$ are written as
\begin{eqnarray}\label{100k}
\fl \Psi_{L,R}(k,t)=G_{L,R}(k,t)-\frac{i}{\pi}\int_{-\infty}^\infty \frac{F_{L,R}(s,t)}{k-s}ds \equiv G_{L,R}(k,t)
-\frac{i}{\pi}\int_{-\infty}^\infty \frac{F_{L,R}(k-s,t)}{ s}ds
\end{eqnarray}
\begin{eqnarray} \label{103}
\fl G_L(k,t)=A_L(k,t)+\frac{1}{2}\left[\widehat{f}_L(k,t)+\widehat{f}_L(-k,t)\right]=
A_L(k,t)\mathcal{T}(k)-F_L(k,t),\nonumber
\\ \fl G_R(k,t)=A_R(-k,t)+\frac{1}{2}\left[\widehat{f}_R(k,t)+\widehat{f}_R(-k,t)\right]=
A_R(-k,t)\mathcal{T}^*(k)+F_R(k,t),\nonumber\\ \fl
F_{L,R}(k,t)=\frac{1}{2}\left[\widehat{f}_{L,R}(k,t)-\widehat{f}_{L,R}(-k,t)\right],\ooo
\widehat{f}_{L,R}(k,t)=A_{L,R}(k,t)\mathcal{R}(k).
\end{eqnarray}
Here we have taken into account the fact that the Fourier transform of the function $\mathrm{sgn}(x)$ is equal to
\begin{eqnarray*}
\fl \widehat{sgn}(k)=-\sqrt{\frac{2}{\pi}}\ooa\frac{i}{k}.
\end{eqnarray*}

We will study the properties of solutions $(\ref{4a})$ and $(\ref{4b})$ using the function $\Psi_L(k,t)$ as an example. We
will assume that $\hbar=1$ and $m=1$, so that $\hbar/(2m)\equiv b=1/2$.

\section{The relationship between the norm of $\Psi_L$ and the function $\mathcal{A}_L(k)$} \label{norma}

Let us express the norm of the solution $\Psi_L(k,t)$ through the function $\mathcal{A}_L(k)$. Taking into account
(\ref{103}), we obtain
\begin{eqnarray}\label{137}
\fl \langle \Psi_L|\Psi_L\rangle= N_1+N_2+N_3=1;
\end{eqnarray}
\begin{eqnarray}\label{138}
\fl N_1=\int_{-\infty}^\infty |G_L(k,t)|^2 dk,\ooo N_2= \frac{1}{\pi}\int_{-\infty}^\infty \int_{-\infty}^\infty \Im
\left[G^*_L(k,t)F_L(q,t)+F_L^*(k,t)G_L(q,t)\right]\frac{dk dq}{k-q},\nonumber \\ \fl N_3=\frac{1}{\pi^2}
\int_{-\infty}^\infty \int_{-\infty}^\infty \int_{-\infty}^\infty \frac{F^*_L(q,t) F_L(s,t)}{(k-q)(k-s)}ds\ooa dq\ooa dk;
\end{eqnarray}
the expression under the integral in $N_2$ is symmetrized with respect to the permutation of $k$ and $q$.

First, we calculate the integral in $N_3$. Taking into account the equality
\begin{eqnarray}\label{788}
\fl \int_{-\infty}^\infty \frac{ds}{s(s-k)}=\pi^2 \delta(k),
\end{eqnarray}
valid for the convolution of the Fourier transform $\widehat{sgn}(k)$, we find
\begin{eqnarray*}
\fl N_3=\frac{1}{\pi^2} \int_{-\infty}^\infty \int_{-\infty}^\infty ds\ooa dq\ooa F^*_L(q,t) F_L(s,t)
\int_{-\infty}^\infty \frac{dk}{(k-q)(k-s)}=\int_{-\infty}^\infty |F_L(q,t)|^2 dq.
\end{eqnarray*}
Finally, taking into account Exps. (\ref{103}), we obtain
\begin{eqnarray}\label{638}
\fl N_1+N_3=\int_{-\infty}^\infty
|A_L(k,t)|^2 dk+\int_{-\infty}^\infty \Re\left[A_L^*(-k,t) \widehat{f}_L(k,t)\right] dk.
\end{eqnarray}

Now, taking into account (\ref{103}), we rewrite the contribution of $N_2$:
\begin{eqnarray}\label{500}
\fl N_2= \frac{1}{2\pi}\int_{-\infty}^\infty \int_{-\infty}^\infty\Im
\bigg[A^*_L(k,t)+\widehat{f}^*_L(k,t)]\widehat{f}_L(q,t)+\widehat{f}^*_L(k,t)A_L(q,t) \nonumber\\
\fl -A^*_L(k,t)\widehat{f}_L(-q,t)-\widehat{f}^*_L(-k,t)A_L(q,t) -\widehat{f}^*_L(-k,t)\widehat{f}_L(-q,t)\bigg]\frac{dk
dq}{k-q}.
\end{eqnarray}
Then, given the expression for $\widehat{f}_L(k,t)$, we make the substitution $k=-k'$ and/or $q =-q'$ in the last three
integrals in (\ref{500}) so that the sign of the arguments of the functions $A_L^*(k,t)$ and $A_L(q,t)$ becomes positive.
As a result, a common factor $A^*_L(k,t)A_L(q,t)$ appears in the integrands in $N_2$:
\begin{eqnarray*}
\fl N_2= \frac{1}{\pi}\int_{-\infty}^\infty \int_{-\infty}^\infty\Im\left\{
A^*_L(k,t)A_L(q,t)\left[(k+q)\mathcal{R}(q)\mathcal{R}^*(k)+k\mathcal{R}^*(k)+q\mathcal{R}(q) \right]\right\}\frac{dk\ooa
dq}{k^2-q^2}.
\end{eqnarray*}
Taking into account (\ref{105}), it is easy to show that the square bracket in this expression is equal to zero.
Therefore, $N_2=0$, and equality (\ref{137}), according to (\ref{638}), takes the form
\begin{eqnarray}\label{133}
\fl \langle \Psi_L|\Psi_L\rangle=  \int_{-\infty}^\infty |\mathcal{A}_L(k)|^2 dk +\int_{-\infty}^\infty
\Re\left[\mathcal{A}^*_L(-k) \mathcal{A}_L(k) \mathcal{R}(k) \right]dk=1.
\end{eqnarray}

As we see, if the function $\mathcal{A}_L(k)$ can take nonzero values on the entire $k$-axis, then the second contribution
in this expression is not equal to zero. Therefore, $\langle \Phi_L^{inc}| \Phi_L^{inc}\rangle=\int_{-\infty}^\infty
|\mathcal{A}_L(k)|^2 dk\neq 1$, and the contribution $\Phi_L^{inc}(x,t)$ in Exp. (\ref{100}) for $\Psi_L(x,t)$ cannot
serve as an in-asymptote of the solution $\Psi_L(x,t)$. But if $\mathcal{A}_L(k)$ is identically zero for $k\leq 0$, as
required by the stationary solution (\ref{3}), then the second contribution is zero and $\langle \Psi_L|\Psi_L\rangle =
\langle \Phi_L^{inc}|\Phi_L^{inc}\rangle=1$. That is, in this case the wave packet $\Phi_L^{inc}(x,t)$ is an in-asymptote
of the solution $\Psi_L(x,t)$.

\section{Asymptotes of the non-stationary solution $\Psi_L$} \label{asymptote}

To find the asymptotes of $\Psi_L(k,t)$ explicitly, we write it down taking into account (\ref{103}):
\begin{eqnarray} \label{300}
\fl \Psi_L(k,t)=\widetilde{G}_L(k) e^{-ibk^2 t}-\frac{i}{2\pi}\int_{-\infty}^\infty
\frac{\widetilde{F}_L(k-s)e^{-ib(k-s)^2 t}-\widetilde{F}_L(k+s)e^{-ib(k+s)^2 t}}{s}ds,
\end{eqnarray}
where $\widetilde{G}_L(k)=G_L(k,0)$, $\widetilde{F}_L(k)=F_L(k,0)$.

We write the sought in- and out-asymptotes in the form $\Psi_L^{in}(k,t)\equiv\widetilde{\Psi}_L^{in}(k)e^ {-ibk^2 t}$ and
$\Psi_L^{out}(k,t)\equiv \widetilde{\Psi}_L^{out}(k) e^{-ibk^2 t}$. They are related to the solution $\Psi_L(k,t)$ by the
equalities
\begin{eqnarray} \label{301}
\fl \widetilde{\Psi}_L^{in}(k)=\lim_{t\to -\infty} e^{ibk^2 t} \Psi_L(k,t);\ppp \widetilde{\Psi}_L^{out}(k)=\lim_{t\to
+\infty} e^{ibk^2 t} \Psi_L(k,t).
\end{eqnarray}

Taking into account Exp. (\ref{300}), we obtain
\begin{eqnarray*}
\fl \widetilde{\Psi}_L^{in/out}=\widetilde{G}_L(k) -\lim_{t\to \mp\infty}\frac{i}{2\pi} \int_{-\infty}^\infty
\left[\widetilde{F}_L(k-s)e^{-ib[(k-s)^2-k^2] t}-\widetilde{F}_L(k+s)e^{-ib[(k+s)^2-k^2] t}\right]\frac{ds}{s}\\ \fl =
\widetilde{G}_L(k) -\lim_{t\to \mp\infty}\frac{i}{2\pi} \int_{-\infty}^\infty \frac{\widetilde{F}_L(k-s)
-\widetilde{F}_L(k+s)}{s} \cos(skt) e^{-ibs^2 t}ds\\ \fl +\lim_{t\to \mp\infty}\frac{1}{2\pi} \int_{-\infty}^\infty
[\widetilde{F}_L(k-s) +\widetilde{F}_L(k+s)] \frac{\sin(skt)}{s} e^{-ibs^2 t}ds.
\end{eqnarray*}
Let's use the stationary phase method. Since
\begin{eqnarray*}
\fl \lim_{t\to \mp\infty} \frac{\sin(skt)}{s}=\mathrm{sgn}(kt)\pi\delta(s),
\end{eqnarray*}
the second integral in $\widetilde{\Psi}_L^{in/out}(k)$ is not equal to zero as $t\to\mp\infty$ due to the contribution of
the stationary phase point $s=0$. The first integral is equal to zero, since its integrand at the points $s=-k$ and $s=k$
is bounded. Taking into account Exps. (\ref{300}) for $\widetilde{G}_L(k)$ and $\widetilde{F}_L(k)$, we obtain
\begin{eqnarray} \label{302}
\fl \widetilde{\Psi}_L^{in}(k)= \widetilde{G}_L(k) -\mathrm{sgn}(k) \widetilde{F}_L(k)= \left\{
\begin{array}{rl}
\mathcal{A}_L(k) \mathcal{T}(k)\ooo (k<0)\\
\mathcal{A}_L(k) +\mathcal{A}_L(-k)\mathcal{R}(-k)\ooo (k>0)
\end{array} \right.;
\end{eqnarray}
\begin{eqnarray*}
\fl \widetilde{\Psi}_L^{out}(k)= \widetilde{G}_L(k) +\mathrm{sgn}(k) \widetilde{F}_L(k))=\left\{
\begin{array}{rl}
\mathcal{A}_L(k) +\mathcal{A}_L(-k)\mathcal{R}(-k)\ooo (k<0)\\ \mathcal{A}_L(k) \mathcal{T}(k)\ooo (k>0)
\end{array} \right. .
\end{eqnarray*}

Taking into account (\ref{105}), it is easy to show that the norms of the asymptotes
$\Psi_L^{in}(k,t)=\widetilde{\Psi}_L^{in}(k)e^{-ibk^2 t}$ and $\Psi_L^{out}(k,t)=\widetilde{\Psi}_L^{out}(k)e^{-ibk^2 t}$
coincide with the norm (\ref{133}) of the solution $\Psi_L$. In particular, it is easy to show that the expression
\begin{eqnarray*}
\fl \int_{-\infty}^\infty \left|\widetilde{\Psi}_L^{in}(k)\right|^2 dk= \int_{-\infty}^0 \left|\mathcal{A}_L(k)
\mathcal{T}(k)\right|^2 dk+ \int_0^\infty  \left|\mathcal{A}_L(k) +\mathcal{A}_L(-k)\mathcal{R}(-k)\right|^2 dk
\end{eqnarray*}
is reduced to Exp. (\ref{133}).

Note that in the coordinate representation, both asymptotes take the form
\begin{eqnarray*}
\fl \Psi_L^{in}(x,t)= \frac{1}{\sqrt{2\pi}}\int_{-\infty}^\infty \Psi_L^{in}(k,t)e^{ikx}dk =
\frac{1}{\sqrt{2\pi}}\int_{-\infty}^0 \mathcal{A}_L(k) \mathcal{T}(k) e^{i(kx-bk^2t)}dk\\ \fl +
\frac{1}{\sqrt{2\pi}}\int_0^\infty \left[\mathcal{A}_L(k)+\mathcal{A}_L(-k) \mathcal{R}(-k)\right]
e^{i(kx-bk^2t)}dk;\nonumber
\end{eqnarray*}
\begin{eqnarray*}
\fl \Psi_L^{out}(x,t)= \frac{1}{\sqrt{2\pi}}\int_{-\infty}^\infty \Psi_L^{out}(k,t)e^{ikx}dk =
\frac{1}{\sqrt{2\pi}}\int_0^\infty \mathcal{A}_L(k) \mathcal{T}(k) e^{i(kx-bk^2t)}dk\\ \fl +
\frac{1}{\sqrt{2\pi}}\int_{-\infty}^0 \left[\mathcal{A}_L(k)+\mathcal{A}_L(-k) \mathcal{R}(-k)\right] e^{i(kx-bk^2t)}dk.
\end{eqnarray*}
With the help of simple transformations they are reduced to the form
\begin{eqnarray} \label{302w}
\fl \Psi_L^{in}(x,t)= \Phi_L^{inc}(x,t) + \Phi_L^{even}(x,t),\ooo \Psi_L^{out}(x,t) = \Phi_L^{tr}(x,t) + \Phi_L^{ref}(x,t)
-\Phi_L^{even}(x,t),
\end{eqnarray}
where
\begin{eqnarray} \label{433}
\fl \Phi_L^{even}(x,t)=\sqrt{\frac{2}{\pi}}\int_0^\infty \mathcal{A}_L(-k) \mathcal{R}(-k)\cos(kx) e^{-ibk^2t}dk.
\end{eqnarray}

As we see, if the function $\mathcal{A}_L(k)$ is non-zero on the entire $k$-axis, then the contribution
$\Phi_L^{inc}(x,t)$ is not an in-asymptote of the solution $\Psi_L(k,t)$, and the superposition of the contributions
$\Phi_L^{tr}(x,t)$ and $\Phi_L^{ref}(x,t)$ is not its out-asymptote, since $\langle
\Phi_L^{inc}|\Phi_L^{inc}\rangle=\langle \Phi_L^{tr}|\Phi_L^{tr}\rangle+\langle \Phi_L^{ref}|\Phi_L^{ref}\rangle\neq
\langle \Psi_L|\Psi_L\rangle$. In this case, the in- and out-asymptotes $\Psi_L^{in}(x,t)$ and $\Psi_L^{out}(x,t)$ of the
solution $\Psi_L(x,t)$ contain a contribution $\Phi_L^{even}(x,t)$ with a time-independent norm, which is a solution of
the Schr\"{o}dinger equation for a free particle, but at the same time the maximum of the modulus $|\Phi_L^{even}(x,t)|$
is achieved at all times at the point $x=0$ where the $\delta$-potential is located (see also the next section). That is,
the dynamics of the non-stationary solution $\Psi_L(x,t)$ is not free as $t\to\mp\infty$, although both of its asymptotes
are solutions of the Schr\"{o}dinger equation for a free particle.

A qualitatively different situation arises when the function $\mathcal{A}_L(k)$ is identically zero for $k\leq 0$. In this
case, the asymptotes $\widetilde{\Psi}_L^{in}(k)$ and $\widetilde{\Psi}_L^{out}(k)$, according to (\ref{302}), are
determined by the expressions
\begin{eqnarray} \label{302a}
\fl \widetilde{\Psi}_L^{in}(k)= \left\{
\begin{array}{rl}
0\ooo (k<0)\\
\mathcal{A}_L(k)\ooo (k>0)
\end{array} \right.;\ppp
\widetilde{\Psi}_L^{out}(k)= \left\{
\begin{array}{rl}
\mathcal{A}_L(-k)\mathcal{R}(-k)\ooo (k<0)\\ \mathcal{A}_L(k) \mathcal{T}(k)\ooo (k>0)
\end{array} \right. .
\end{eqnarray}
Similarly, when $\mathcal{A}_R(k)$ is identically equal to zero for $k\leq 0$, then the asymptotes of the solution
$\Psi_R(k,t)$ are determined by the expressions
\begin{eqnarray} \label{302b}
\fl \widetilde{\Psi}_R^{in}(k)= \left\{
\begin{array}{rl}
\mathcal{A}_R(-k) \ooo (k<0) \\
0\ooo (k>0)
\end{array} \right.;\ppp
\widetilde{\Psi}_R^{out}(k)= \left\{
\begin{array}{rl}
\mathcal{A}_R(-k) \mathcal{T}(-k)\ooo (k<0) \\ \mathcal{A}_R(k)\mathcal{R}(k)\ooo (k>0)
\end{array} \right. .
\end{eqnarray}

As we see, for this class of functions $\mathcal{A}_L(k)$ and $\mathcal{A}_R(k)$, the in-asymptotes of the solutions
$\Psi_L(k,t)$ and $\Psi_R(k,t)$ are the wave packets $\widehat{\Phi}_L^{inc}(k,t)$ and $\widehat{\Phi}_R^{inc}(k,t)$, and
the out-asymptote of each of them is a superposition of the reflected and transmitted wave packets:
\begin{eqnarray*}
\fl \Psi_{L,R}^{in}(k,t)=\widehat{\Phi}_{L,R}^{inc}(k,t),\ppp
\Psi_{L,R}^{out}(k,t)=\widehat{\Phi}_{L,R}^{ref}(k,t)+\widehat{\Phi}_{L,R}^{tr}(k,t);
\\
\fl \widehat{\Phi}_L^{inc}(k,t)= \mathcal{A}_L(k)e^{-ibk^2 t} \in \widehat{\mathcal{S}}_{(+)},\ppp\ppp\ppp\ppp
\widehat{\Phi}_R^{inc}(k,t)= \mathcal{A}_R(-k)e^{-ibk^2 t} \in \widehat{\mathcal{S}}_{(-)}, \\ \fl
\widehat{\Phi}_L^{ref}(k,t)= \mathcal{A}_L(-k)\mathcal{R}(-k) e^{-ibk^2 t} \in \widehat{\mathcal{S}}_{(-)},  \ooo\ooa
\widehat{\Phi}_R^{ref}(k,t)= \mathcal{A}_R(k)\mathcal{R}(k) e^{-ibk^2 t} \in \widehat{\mathcal{S}}_{(+)},\\ \fl
\widehat{\Phi}_L^{tr}(k,t)= \mathcal{A}_L(k)\mathcal{T}(k) e^{-ibk^2 t} \in \widehat{\mathcal{S}}_{(+)},\ppp\ppp\ooo
\widehat{\Phi}_R^{tr}(k,t)= \mathcal{A}_R(-k)\mathcal{T}(-k) e^{-ibk^2 t} \in \widehat{\mathcal{S}}_{(-)};
\end{eqnarray*}
where $\widehat{\mathcal{S}}_{(+)}=\mathcal{S}(\widehat{\mathbb{R}})\bigcap L^2(\widehat{\mathbb{R}}_+)$,
$\widehat{\mathcal{S}}_{(-)}=\mathcal{S}(\widehat{\mathbb{R}})\bigcap L^2(\widehat{\mathbb{R}}_-)$;
$\mathcal{S}(\widehat{\mathbb{R}})$ -- Schwartz space in $k$-space; $\widehat{\mathbb{R}}_-$ and $\widehat{\mathbb{R}}_+$
are, respectively, the intervals $(-\infty,0)$ and $(0.\infty)$ in $k$-space.

Thus, if the function $\mathcal{A}_L(k)$ is not equal to zero on the entire $k$-axis, then the norms
$\langle\Phi_L^{tr}|\Phi_L^{tr}\rangle$ and $\langle\Phi_L^{ref}|\Phi_L^{ref}\rangle$ cannot be interpreted as
transmission and reflection coefficients, since $\langle\Phi_L^{tr}|\Phi_L^{tr}\rangle+\langle
\Phi_L^{ref}|\Phi_L^{ref}\rangle\neq 1$ (although $|\mathcal{T}(k)|^2+|\mathcal{R}(k)|^2=1$ for all values of $k$). In
this case, the solution $\Psi_L(k,t)$ defines a pure state, but it describes not the process of {\it scattering} of a
particle on a NDP, but the process of its {\it interaction} with this potential, which lasts infinitely long. If the
function $\mathcal{A}_L(k)$ is identically zero for $k\leq 0$, then the norms $\langle\Phi_L^{tr}|\Phi_L^{tr}\rangle$ and
$\langle\Phi_L^{ref}|\Phi_L^{ref}\rangle$ are the transmission and reflection coefficients, since $\langle
\Phi_L^{tr}|\Phi_L^{tr}\rangle+\langle \Phi_L^{ref}|\Phi_L^{ref}\rangle= 1$. In this case, as will be shown below, the
solution $\Psi_L(k,t)$ defines the scattering state. A similar situation arises for the solution $\Psi_R(k,t)$.

The main result of this section is that the space of states of the particle in the problem under study is the sum of two
subspaces: $\mathcal{H}=\mathcal{H}_{ac}=\mathcal{H}_{non-scat}\oplus\mathcal{H}_{scat}$, where $\mathcal{H}_{non-scat}$
is the subspace of pure states of the particle, which have no free dynamics as $t\to\mp\infty$, and $\mathcal{H}_{scat}$
is the subspace of scattering states, the study of whose properties is the subject of all subsequent sections of this
paper.

\section{On the absence of transitions between the subspaces $\widehat{\mathcal{S}}_{(+)}$ and $\widehat{\mathcal{S}}_{(-)}$
under the action of observables' operators} \label{invar}

Consider the scattering problem with the two-sided incidence of a particle onto the barrier, which is described by a
superposition of the solutions $\Psi_L(x,t)$ and $\Psi_R(x,t)$ obtained using the functions $\mathcal{A}_L(k)$ and
$\mathcal{A}_R(k)$ from $\mathcal{\widehat{S}}_{(+)}$. We need to calculate the scalar products
$\langle\Psi_L|\hat{x}^n|\Psi_R\rangle_t$ and $\langle\Psi_L|\hat{p}^n|\Psi_R\rangle_t$ in the $k$-representation as $t\to
-\infty$ for $n=0,1,\ldots$

According to (\ref{100}), in the $x$-representation
\begin{eqnarray*} \label{141}
\fl \hat{x}^n\Psi_{R}(x,t)= x^n\left\{\Phi_{R}^{inc}(x,t)+\frac{1}{2}\left[f_{R}(x,t)+f_{R}(-x,t)\right]\right\}
+\frac{\mathrm{sgn}(x)}{2}x^n\left[f_{R}(x,t)-f_{R}(-x,t)\right]\\
\fl \hat{k}^n\Psi_{R}(x,t)= (-i)^n\frac{\partial^n}{\partial
x^n}\left\{\Phi_{R}^{inc}(x,t)+\frac{1}{2}\left[f_{R}(x,t)+f_{R}(-x,t)\right]\right\} \nonumber \\ \fl +(-i)^n
\frac{\mathrm{sgn}(x)}{2}\frac{\partial^n}{\partial x^n}\left[f_{R}(x,t)-f_{R}(-x,t)\right].
\end{eqnarray*}
Thus, in the $k$-representation
\begin{eqnarray*}
\fl \widehat{(\hat{x}^n\Psi_R)}(k,t)=i^n\left[\frac{\partial^n G_R(k,t)}{\partial k^n}-\frac{i}{\pi}\int_{-\infty}^\infty
\frac{\partial^n F_R(s,t)}{\partial s^n} \frac{ ds}{k-s}\right];\\ \fl \widehat{(\hat{k}^n\Psi_R)}(k,t)=k^n
G_R(k,t)-\frac{i}{\pi}\int_{-\infty}^\infty s^n F_R(s,t)\frac{ ds}{k-s}.
\end{eqnarray*}

Since we are interested in both scalar products only for $t\to -\infty$, when calculating
$\widehat{(\hat{x}^n\Psi_R)}(k,t)$ it is sufficient to take into account only those terms that contain the derivative
$\frac{\partial^n }{dk^n} e^{-ibk^2t}=(-ikt)^n e^{-ibk^2t}$:
\begin{eqnarray*}\label{140a}
\fl \lim_{t\to -\infty}\widehat{(\hat{x}^n\Psi_R)}(k,t)=t^n\left[k^n G_R(k,t)-\frac{i}{\pi}\int_{-\infty}^\infty \frac{s^n
F_R(s,t)}{k-s}ds\right]=t^n \widehat{(\hat{k}^n\Psi_R)}(k,t).
\end{eqnarray*}
Thus, calculating the scalar product $\langle\Psi_L|\hat{x}^n|\Psi_R\rangle_t$ as $t\to -\infty$ reduces to calculating
$\langle\Psi_L|\hat{k}^n|\Psi_R\rangle_t$.

Similarly to Exps. (\ref{137}) and (\ref{138}), for $\langle\Psi_L|\hat{k}^n|\Psi_R\rangle_t$ we have
\begin{eqnarray}\label{147}
\fl \langle\Psi_L|\hat{k}^n|\Psi_R\rangle_t= K_1+K_2+K_3;
\end{eqnarray}
\begin{eqnarray}\label{148}
\fl K_1=\int_{-\infty}^\infty k^n G_L^*(k,t)G_R(k,t) dk,\ooo K_3=\frac{1}{\pi^2} \int_{-\infty}^\infty
\int_{-\infty}^\infty \int_{-\infty}^\infty \frac{ s^n F_L^*(q,t)F_R(s,t)}{(k-q)(k-s)}ds\ooa dq\ooa dk,\nonumber\\ \fl
K_2= \frac{i}{2\pi} \int_{-\infty}^\infty \int_{-\infty}^\infty \Big\{k^n \left[F_L^*(q,t)G_R(k,t)+
G_L^*(q,t)F_R(k,t)\right]\nonumber\\ \fl -q^n\left[G_L^*(k,t)F_R(q,t)+F_L^*(k,t) G_R(q,t)\right]\Big\}\frac{dk dq}{k-q};
\end{eqnarray}
As in Exp. (\ref{138}) for $N_2$, the integrand in $K_2$ is symmetrized with respect to the permutation of the integration
variables $k$ and $q$.

Taking into account (\ref{788}), we calculate $K_3$ --
\begin{eqnarray*}
\fl K_3=\frac{1}{\pi^2} \int_{-\infty}^\infty \int_{-\infty}^\infty ds\ooa dq\ooa s^n F_L^*(q,t) F_R(s,t)
\int_{-\infty}^\infty \frac{dk}{(k-q)(k-s)}=\int_{-\infty}^\infty q^n F_L^*(q,t)F_R(q,t) dq
\end{eqnarray*}
Next, taking into account the identity $\mathcal{A}_{L,R}(k)\mathcal{A}_{L,R}(-k)\equiv 0$, as well as Exps. (\ref{105})
and (\ref{103}), we find $K_1+K_3$:
\begin{eqnarray}\label{150}
\fl K_1+K_3= \chi_n\int_0^\infty k^n \mathcal{A}_L^*(k)\mathcal{A}_R(k)\mathcal{T}^*(k)\mathcal{R}(k) dk; \ppp
\chi_n=\frac{1-(-1)^n}{2}.
\end{eqnarray}

As for $K_2$ in (\ref{148}), taking into account (\ref{103}), it is easy to show that the expression in the first square
bracket reduces to the form
\begin{eqnarray*}
\fl F_L^*(q,t)G_R(k,t)+ G_L^*(q,t)F_R(k,t)=\frac{1}{2}\bigg[A^*_L(q,t)A_R(-k,t)[R^*(q)-R^*(k)]\\
\fl +A^*_L(q,t)A_R(k,t)R(k)T^*(q)-A^*_L(-q,t)A_R(-k,t)R(q)T^*(k)\bigg];
\end{eqnarray*}
If we swap $k$ and $q$, we obtain the second square bracket in $K_2$ (see (\ref{148})).

Next, by analogy with $N_2$, we transform the resulting expressions so that the sign of the arguments of functions $A_L$
and $A_R$ is positive. We also take into account Exps. (\ref{4a}) and (\ref{4b}) for these functions. As a result, $K_2$
is reduced to the form
\begin{eqnarray*}
\fl K_2= \int_0^\infty \int_0^\infty \Big[ [\rho(k,q)+\rho(q,k)] \frac{\sin bt(k^2-q^2)}{k^2-q^2}+
i\frac{\rho(k,q)-\rho(q,k)}{k^2-q^2} \cos bt(k^2-q^2) \Big] \frac{dq\ooa dk}{4\pi}
\end{eqnarray*}
\begin{eqnarray*}
\fl \rho(k,q)= \Big[(-1)^n (k-q) [\mathcal{R}(k)-\mathcal{R}^*(q)]
+(k+q)\left[(-1)^n\mathcal{R}^*(q)\mathcal{T}(k)+\mathcal{R}(k)\mathcal{T}^*(q)\right]\Big]\\ \fl \times
k^n\mathcal{A}^*_L(q)\mathcal{A}_\mathcal{R}(k).
\end{eqnarray*}

It is then convenient to move from the pair of variables $(k,q)$ to the variables $(y,z)$: $k=z+y$, $q=z-y$; the Jacobian
of the transformation is equal to two. Since $k>0$ and $q>0$, the new variables change in the regions $z\geq 0$ and
$|y|\leq z$. As a result, we obtain
\begin{eqnarray}\label{152}
\fl K_2= \frac{1}{8\pi}\int_0^\infty \frac{dz}{z} \int_{-z}^z dy \bigg\{\bigg[\rho(z+y,z-y)+\rho(z-y,z+y)\bigg]
\frac{\sin(2yzt)}{y}\nonumber \\
\fl + i\frac{\rho(z+y,z-y)-\rho(z-y,z+y)}{y} \cos(2yzt)\bigg\}.
\end{eqnarray}

According to the stationary phase method, as $t\to -\infty$, the integrals over $y$ of expressions containing rapidly
oscillating functions $\sin(2yzt)$ and $\cos(2yzt)$ can differ from zero only due to the contribution of the stationary
point $y=0$. Obviously, the integral over $y$ of an expression containing $\cos(2yzt)$ is zero, since this expression is a
bounded function at this point. However, the integral of an expression containing $\sin(2yzt)$ is not zero, since
$\lim_{t\to -\infty}\frac{\sin(2yzt)}{y}= -\pi\delta(y)$:
\begin{eqnarray*}
\fl \lim_{t\to -\infty} K_2=- \int_0^\infty \frac{\rho(z,z)}{4z}dz = -\chi_n \int_0^\infty z^n
\mathcal{A}^*_L(z)\mathcal{A}_R(z)\mathcal{R}(z)\mathcal{T}^*(z).
\end{eqnarray*}
That is, when $t\to-\infty$
\begin{eqnarray}\label{154}
\fl \langle\Psi_L|\hat{k}^n|\Psi_R\rangle_t =K_2+(K_1+K_3)=0.
\end{eqnarray}
As a consequence, the scalar product $\langle\Psi_L|\hat{x}^n|\Psi_R\rangle_t$ for any value $n\geq 0$ is also zero in
this limit.

Thus, if ${A}_L(k), {A}_R(k)\in\widehat{\mathcal{S}}_{(+)}$, then as $t\to-\infty$, when
$\Psi_L(k,t)\in\widehat{\mathcal{S}}_{(+)}$, and $\Psi_R(k,t)\in\widehat{\mathcal{S}}_{(-)}$ (see section
\ref{asymptote}), the equalities $\langle\Psi_L|\hat{x}^n|\Psi_R\rangle_t=0$ and
$\langle\Psi_L|\hat{k}^n|\Psi_R\rangle_t=0$ hold. And since for a spinless particle the operator of an arbitrary
observable $O$ is a polynomial in $\hat{x}$ and $\hat{p}=\hbar\hat{k}$, then the previous two equalities can be combined
into one: $\langle\Psi_L|\hat{O}|\Psi_R\rangle_t=0$.

Similar equalities arise for each of the solutions $\Psi_L(k,t)$ and $\Psi_R(k,t)$ as $t\to +\infty$, since each of them
is in this case a superposition of the transmitted and reflected wave packets -- out-asymptotes from the subspaces
$\widehat{\mathcal{S}}_{(+)}$ and $\widehat{\mathcal{S}}_{(-)}$ (see ibid.); now
$\langle\widehat{\Phi}^{tr}_{L,R}|\hat{O}|\widehat{\Phi}^{ref}_{L,R}\rangle_t=0$.

That is, if ${A}_L(k), {A}_R(k)\in\widehat{\mathcal{S}}_{(+)}$, then in the $k$-representation, the non-stationary
solution of general form is, as $t\to\mp\infty$, a superposition of asymptotes from the subspaces
$\widehat{\mathcal{S}}_{(+)}$ and $\widehat{\mathcal{S}}_{(-)}$, invariant with respect to the action of the observables'
operators. This remains true in the coordinate representation as well, since the values of the scalar products do not
depend on the choice of representation. But now, instead of the subspaces $\widehat{\mathcal{S}}_{(+)}$ and
$\widehat{\mathcal{S}}_{(-)}$, which are characterized by the {\it signs of the momentum} of the particle at
$t\to\mp\infty$, we are dealing with the functional subspace of {\it left} asymptotes $\mathcal{S}_{(-)}=
\mathcal{S}({\mathbb{R}}_-)$ and the subspace of {\it right} asymptotes $\mathcal{S}_{(+)}=\mathcal{S}({\mathbb{R}}_+)$,
localized, respectively, to the left and right of the barrier. In particular, if $t\to -\infty$, then $\Psi_L(x,t)\in
\mathcal{S}_{(-)}$, and $\Psi_R(x,t)\in \mathcal{S}_{(+)}$.

Thus, in the $k$-representation there are no transitions under the action of observable operators between the subspaces
$\widehat{\mathcal{S}}_{(+)}$ and $\widehat{\mathcal{S}}_{(-)}$ formed by the asymptotes with $k>0$ and $k<0$, and in the
$x$-representation there are no transitions between the subspaces ${\mathcal{S}}_{(+)}$ and ${\mathcal{S}}_{(-)}$ formed
by the right and left asymptotes. This means that the asymptotes from these subspaces do not interact with each other and
describe the free dynamics as $t\to +\infty$. In other words, the non-stationary solutions $\Psi_L(k,t)$ and
$\Psi_R(k,t)$, constructed using the functions ${A}_L(k)$ and ${A}_R(k)$ from $\widehat{\mathcal{S}}_{(+)}$, determine the
scattering states. That is, it is these solutions that form the subspace $\mathcal{H}_{scat}$ introduced at the end of the
previous section.

\section{Asymptotic superselection rule} \label{rule}

Thus, as $t\to\mp\infty$, the scattering state space $\mathcal{H}_{scat}$ is a direct sum of two subspaces that are
invariant with respect to the action of the observables' operators: in the $x$-representation, the space of state vectors
$\mathcal{H}_{scat}$ is represented as $t\to\mp\infty$ by the functional space
$\mathcal{S}_{scat}=\mathcal{S}_{(-)}\oplus\mathcal{S}_{(+)}$, and in the $k$-representation, by the space
$\widehat{\mathcal{S}}_{scat}=\widehat{\mathcal{S}}_{(-)} \oplus\widehat{\mathcal{S}}_{(+)}$. This means that the
operators of the observables act in $\mathcal{H}_{scat}$ {\it reducibly}: they are defined, as linear operators, not on
the entire $\mathcal{H}_{scat}$, but only in its invariant subspaces. According to the theory of superselection rules
(see, e.g., \cite{Hor1,Hor2,Hor3,Ear}), this property means that the theory of this scattering process is based on a
superselection rule. And since superselection arises here as $t\to\mp\infty$, we are talking about an {\it asymptotic}
superselection rule.

To study the nature of this superselection rule in more detail, consider in the k-representation as $t\to \mp\infty$ the
family of asymptotes $\Phi=\alpha\Phi_{(-)}+\beta\Phi_{(+)}$, where $\Phi_{(+) }\in \mathcal{\widehat{S}}_{(+)}$,
$\Phi_{(-)}\in \mathcal{\widehat{S}}_{(-)}$, and $|\alpha|^2+|\beta|^2=1$. As shown in Section \ref{invar}, the states in
these subspaces satisfy the equality
\begin{eqnarray}\label{176}
\fl \langle\Phi_{(-)}|\hat{O}|\Phi_{(+)}\rangle=0
\end{eqnarray}
for any observable $O$. Given this equality, it is easy to show that the average value of this observable for a particle
in state $\Phi$ is defined as
\begin{eqnarray}\label{177}
\fl \langle\Phi|\hat{O}|\Phi\rangle= |\alpha|^2\langle\Phi_{(-)}|\hat{O}|\Phi_{(-)}\rangle+
|\beta|^2\langle\Phi_{(+)}|\hat{O}|\Phi_{(+)}\rangle.
\end{eqnarray}
That is, due to the property (\ref{176}), there are no interference terms in this expression. This means that the
superposition $|\Phi\rangle=\alpha|\Phi_{(-)}\rangle+\beta|\Phi_{(+)}\rangle$ describes a mixed state, although it is a
solution to the Schr\"{o}dinger equation.

According to \cite{Hor1,Hor2,Hor3}, in addition to the two interrelated conditions (\ref{176}) and (\ref{177}), the
existence condition of a superselection operator should also be viewed as a condition for the existence of a
superselection rule in the theory. This operator commutes with the operators of all observables, and the subspaces
$\mathcal{\widehat{S}}_{(+)}$ and $\mathcal{\widehat{S}}_{(-)}$ (or $\mathcal{S}_{(+)}$ and $\mathcal{S}_{(-)}$), called
superselection sectors, are its proper subspaces.

The role of the superselection operator in the space
$\mathcal{\widehat{S}}_{scat}=\mathcal{\widehat{S}}_{(-)}\oplus\mathcal{\widehat{S}}_{(+)}$ as $t\to \mp\infty$ is played
by the operator $\hat{P}_k=\mathrm{sgn}(k)$. Obviously, for the functions $\Phi_{(-)}(k,t)\in\mathcal{\widehat{S}}_{(-)}$
and $\Phi_{(+)}(k,t)\in \mathcal{\widehat{S}}_{(+)}$
\begin{eqnarray*}
\fl\hat{P}_k\Phi_{(-)}=-\Phi_{(-)},\ooo \hat{P}_k\Phi_{(+)}=+\Phi_{(+)}.
\end{eqnarray*}
And since $\Phi_{(-)}(k,t)$ and $\Phi_{(+)}(k,t)$ are arbitrary functions from $\mathcal{\widehat{S}}_{(-)}$ and
$\mathcal{\widehat{S}}_{(+)}$, these subspaces are eigenspaces of the operator $\hat{P}_k$. Moreover, it is important to
emphasize that in the open intervals $\widehat{\mathbb{R}}_-$ and $\widehat{\mathbb{R}}_+$, where functions from the
sectors $\mathcal{\widehat{S}}_{(-)}$ and $\mathcal{\widehat{S}}_{(+)}$ are defined, the operator $\hat{P}_k$ commutes
with the operators $\hat{x}^n$ and $\hat{p}^n$ for any natural $n$.

We will show that in the $x$-representation the superselection operator is the operator $\hat{P}_{x}=\mathrm{sgn}(x)$, and
the asymptotes from the eigen spaces (coherent sectors) $\mathcal{\widehat{S}}_{(-)}$ and $\mathcal{\widehat{S}}_{(+)}$ of
the operator $\hat{P}_{k}$ are also eigen spaces of the operator $\hat{P}_{x}$. In other words, we need to show that in
this scattering problem the asymptotic superselection rule is combined in the sense that, in addition to the selection by
the sign of the particle momentum at $t\to\mp\infty$, in these two limiting cases there is also a selection of states by
the region of localization of the particle state (located to the left or to the right of the barrier). Let's do this using
the example of the asymptote $\Phi_{(+)}(k,t)=\mathcal{A}(k)e^{-ibk^2t}\in \mathcal{\widehat{S}}_{(+)}$, for which
$\hat{P}_{k}|\Phi_{(+)}\rangle =+|\Phi_{(+)}\rangle$. Our task is to calculate the result of the action of the operator
$\hat{P}_{x}$ on this state.

Since the Fourier transform of the function $\mathrm{sgn}(x)$ is $-\sqrt{\frac{2}{\pi}}\ooa\frac{i}{k}$, the state
$|\widetilde{\Phi}_{(+)}\rangle=\hat{P}_{x}|\Phi_{(+)}\rangle$ in the $k$-representation can be written as
\begin{eqnarray*}
\fl \widetilde{\Phi}_{(+)}(k,t)=\frac{i}{2\pi}\int_{-\infty}^\infty \left[\mathcal{A}(k+s)e^{-ib(k+s)^2
t}-\mathcal{A}(k-s)e^{-ib(k-s)^2 t}\right]\frac{ds}{s}.
\end{eqnarray*}
And since we are interested in this state only for $t\to\mp\infty$, it suffices to find its in- and out-asymptotes
$\widetilde{\Phi}_{(+)}^{in,out}(k,t)$. Proceeding in the same way as in the analysis of the asymptotes of the state
$\Psi_L(k,t)$ (see section \ref{asymptote}), we find
\begin{eqnarray*}
\fl \widetilde{\Phi}_{(+)}^{in}(k,t)=-\mathrm{sgn}(k) \mathcal{A}(k)e^{-ibk^2 t}=-\Phi_{(+)}(k,t)\ooo\mbox{for}\ooo t\to -\infty;\\
\fl\widetilde{\Phi}_{(+)}^{out}(k,t)=+\mathrm{sgn}(k) \mathcal{A}(k)e^{-ibk^2 t}=+\Phi_{(+)}(k,t)\ooo\mbox{for}\ooo  t\to
+\infty
\end{eqnarray*}
(Here we have taken into account the fact that $\Phi_{(+)}(k,t)\equiv 0$ for $k< 0$). That is,
\begin{eqnarray*}
\fl\hat{P}_{x}|\Phi_{(+)}\rangle=-|\Phi_{(+)}\rangle\ooo\mbox{for}\ooo  t\to -\infty;\ppp
\hat{P}_{x}|\Phi_{(+)}\rangle=+|\Phi_{(+)}\rangle\ooo\mbox{for}\ooo t\to +\infty.
\end{eqnarray*}
Q.E.D.

Similarly, for the state $\Phi_{(-)}(k,t)=\mathcal{A}(-k) e^{-ibk^2t}$ from $\mathcal{\widehat{S} }_{(-)}$, for which
$\hat{P}_{k}|\Phi_{(-)}\rangle=-|\Phi_{(-)}\rangle$, we find
\begin{eqnarray*}
\fl \hat{P}_{x}|\Phi_{(-)}\rangle=+|\Phi_{(-)}\rangle\ooo \mbox{lkz}\ooo t\to -\infty;\ppp
\hat{P}_{x}|\Phi_{(-)}\rangle=-|\Phi_{(-)}\rangle\ooo\mbox{for}\ooo t\to +\infty.
\end{eqnarray*}
All this means that in the space of asymptotes as $t\to \mp\infty$, the eigensubspaces of the superselection operators
$\hat{P}_{k}$ and $\hat{P}_{x}$ coincide and, consequently, in these subspaces (superselection sectors) they commute with
each other.

Thus, as $t\to \mp\infty$, the asymptotes from $\mathcal{\widehat{S}}_{scat}=
\mathcal{\widehat{S}}_{(-)}\oplus\mathcal{\widehat{S}}_{(+)}$ in the $k$-representation form in the $x$-representation the
space $\mathcal{S}_{scat}=\mathcal{S}_{(-)}\oplus\mathcal{S}_{(+)}$, where the subspace $\mathcal{S}_{(-)}$ is formed by
the left asymptotes (that is, the asymptotes localized, as $t\to \mp\infty$, to the left of the barrier), and the subspace
$\mathcal{S}_{(+)}$ is formed by the right asymptotes (that is, the asymptotes localized, when $t\to \mp\infty$, to the
right of the barrier). In other words, left and right asymptotes form superselection sectors in the space of asymptotes in
the $x$-representation as $t\to \mp\infty$.

Obviously, one must distinguish between left and right asymptotes whose Fourier-images belong to the subspace
$\mathcal{\widehat{S}}_{(-)}$ in the $k$-representation, and left and right asymptotes whose Fourier-images belong to the
subspace $\mathcal{\widehat{S}}_{(+)}$. Thus, in this problem, four different superselection sectors arise in the space of
{\it vectors} of scattering states: two at $t\to -\infty$ and two at $t\to +\infty$. This means that in this scattering
problem, the space of vectors of scattering states $\mathcal{H}_{scat}$ depends on time and should be denoted as
$\mathcal{H}_{scat}(t)$.

And since the functional subspaces $\mathcal{S}_{(-)}$ and $\mathcal{S}_{(+)}$ correspond to the eigenvalues $-1$ and $+1$
of the superselection operator $\mathrm{sgn}(x)$, and the subspaces $\mathcal{\widehat{S}}_{(-)}$ and
$\mathcal{\widehat{S}}_{(+)}$ correspond to the eigenvalues $-1$ and $+1$ of the superselection operator
$\mathrm{sgn}(k)$, then each of the four superselection sectors in $\mathcal{H}_{scat}(t)$ as $t\to \mp\infty$ is
conveniently denoted as $\mathcal{H}_{(\mathrm{sgn}(x),\mathrm{sgn}(k))}$:
\begin{itemize}
\item[(1)] $\mathcal{H}_{(-1,+1)}$ -- sector of left in-asymptotes;
\item[(2)] $\mathcal{H}_{(+1,-1)}$ -- sector of right in-asymptotes;
\item[(3)] $\mathcal{H}_{(-1,-1)}$ -- sector of left out-asymptotes;
\item[(4)] $\mathcal{H}_{(+1,+1)}$ -- sector of right out-asymptotes.
\end{itemize}

Thus,
\begin{eqnarray}\label{1000}
\fl \lim_{t\to -\infty}\mathcal{H}_{scat}(t)=\mathcal{H}_{(-1,+1)}\oplus \mathcal{H}_{(+1,-1)};\ppp \lim_{t\to
+\infty}\mathcal{H}_{scat}(t)=\mathcal{H}_{(-1,-1)}\oplus \mathcal{H}_{(+1,+1)}.
\end{eqnarray}
At intermediate moments of time, the Schr\"{o}dinger dynamics crosses the boundaries of superselection sectors in
$\mathcal{H}_{scat}(t)$.

According to the asymptotic superselection rule valid in this problem, as $t\to\mp\infty$, any superposition of asymptotes
from the same superselected sector $\mathcal{H}_{scat}(t)$ is a pure state, and any superposition of asymptotes from
different sectors is a mixed state. The operators of observables, including the Hamiltonian $H$, are defined only in the
superselection sectors. This is effectively a new formulation of the superposition principle for this scattering problem

For example, in the scattering problem with the left-sided incidence of a particle onto the barrier, the particle's state
as $t\to -\infty$ is a pure state from the sector $\mathcal{H}_{(-1,+1)}$, while at $t\to +\infty$ it turns out to be a
mixed state, which is a superposition of a pure state from the sector $\mathcal{H}_{(-1,-1)}$ and a pure state from the
sector $\mathcal{H}_{(+1,+1)}$. That is, this scattering process is actually a decoherence process in a closed system
(this also applies to the scattering process with a particle falling to the right on a barrier).

Since the operators of observables are defined only in superselection sectors, the scattering process with one-sided
particle incidence on the NDP must be considered as a mixture of two subprocesses: the subprocess of particle tunneling
(passage) through the NDP and the subprocess of particle reflection from the NDP. Consequently, the model of this
scattering process must provide a description of these subprocesses at all stages of scattering. And since the division
into subprocesses at $t\to -\infty$ arises during Schr\"{o}dinger dynamics itself, the main task is to describe the
subprocesses at preceding stages of scattering, including as $t\to -\infty$.

For example, for the state $\Psi_{L}(x,k)$, which describes the scattering process with a left-sided incidence of a
particle on the barrier, this can be done as follows. We rewrite the stationary state $\Psi_{L}(x,k)$ as
\begin{eqnarray*}
\fl \Psi_{L}(x,k)= \mathcal{T}(k) e^{ikx}-\theta(-x)\mathcal{R}(k)\left(e^{ikx}- e^{-ikx}\right).
\end{eqnarray*}
As we can see, this function $\Psi_{L}(x,t)$ is a superposition
\begin{eqnarray} \label{183}
\fl \Psi_{L}(x,t)= \Psi_{tr}(x,t)+\Psi_{ref}(x,t),
\end{eqnarray}
where the wave packet $\Psi_{tr}(x,t)=\int_{-\infty}^\infty A_L(k,t)\mathcal{T}(k) e^{ikx}dk$ describes the transmission
subprocess, and $\Psi_{ref}(x,t)=-2i\theta(-x)\int_{-\infty}^\infty A_L(k,t)\mathcal{R}(k)\sin(kx)dk$ is the reflection
subprocess. As $t\to -\infty$, both subprocesses are described by wave functions from the superselection sector
$\mathcal{H}_{(-1,+1)}$. For $t\to +\infty$, the transmission subprocess is described by a wave function from
$\mathcal{H}_{(+1,+1)}$, and the reflection subprocess is described by a wave function from $\mathcal{H}_{(-1,-1)}$.

\section{Conclusion} \label{Conclude}

Thus, there are two quantum-mechanical models of particle scattering on a one-dimensional $\delta$-potential -- the
standard model, based on the concept of wave operators, and a new model, based on the analysis of asymptotes of
non-stationary solutions of the Schr\"{o}dinger equation as $t\to\mp\infty$. According to the concept of wave operators,
the entire space of states of the absolutely continuous spectrum $\mathcal{H}_{ac}$ in this problem consists of scattering
states, which are pure states. According to the Schr\"{o}dinger equation, this space is the sum of a subspace of pure
states that have no free dynamics as $t\to\mp\infty$, and a subspace of states that have asymptotically free dynamics, but
are mixed states. That is, according to the Schr\"{o}dinger equation, in this scattering problem, assumptions (a) and (b),
underlying the concept of wave operators, exclude each other.

\end{document}